\documentclass[twocolumn,reprint,aps,prd,nofootinbib,floatfix,superscriptaddress]{revtex4-1}
\usepackage[utf8]{inputenc}
\usepackage[utf8]{inputenc}
\usepackage{amsmath}
\usepackage{amsfonts}
\usepackage{amssymb}
\usepackage{float}
\usepackage[left=2cm,right=2cm,top=2cm,bottom=2cm]{geometry}
\usepackage{braket}
\usepackage{dsfont}
\usepackage{hyperref}

\usepackage{graphicx}
\usepackage{tikz}
\usetikzlibrary{arrows,decorations.pathmorphing,backgrounds,positioning,fit,petri,shapes}
\usepackage{lipsum}

\newcommand\blfootnote[1]{%
  \begingroup
  \renewcommand\thefootnote{}\footnote{#1}%
  \addtocounter{footnote}{-1}%
  \endgroup
}

\usepackage{multirow}
\begin{document}

\title{Constraining $f(R)$ gravity with a $k$-cut cosmic shear analysis of the Hyper Suprime-Cam first-year data}

\begin{abstract}

Using Subaru Hyper Suprime-Cam (HSC) year 1 data, we perform the first $k$-cut cosmic shear analysis constraining both $\Lambda$CDM and $f(R)$ Hu-Sawicki modified gravity. To generate the $f(R)$ cosmic shear theory vector, we use the matter power spectrum emulator trained on COLA (COmoving Lagrangian Acceleration) simulations \cite{Ramachandra:2020lue}. The $k$-cut method is used to significantly down-weight sensitivity to small scale ($k > 1 \ h {\rm Mpc }^{-1}$) modes in the matter power spectrum where the emulator is less accurate, while simultaneously ensuring our results are robust to baryonic feedback model uncertainty. We have also developed a test to ensure that the effects of poorly modeled small scales are nulled as intended. For $\Lambda$CDM we find $S_8 = \sigma_8 (\Omega_m / 0.3) ^ {0.5} = 0.789 ^{+0.039}_{-0.022}$, while the constraints on the $f(R)$ modified gravity parameters are prior dominated. In the future, the $k$-cut  method could be used to constrain a large number of theories of gravity where computational limitations make it infeasible to model the matter power spectrum down to extremely small scales.

\end{abstract}

\author{Leah Vazsonyi}
\email{lvazsony@caltech.edu}
\blfootnote{© 2021. All rights reserved.} 
\affiliation{California Institute of Technology, 1201 East California Blvd., Pasadena, California 91125, USA}
\affiliation{Jet Propulsion Laboratory, California Institute of Technology, 4800 Oak Grove Drive, Pasadena, California 91109, USA}
\author{Peter L.~Taylor}
\affiliation{Jet Propulsion Laboratory, California Institute of Technology, 4800 Oak Grove Drive, Pasadena, California 91109, USA}
\author{Georgios Valogiannis}
\affiliation{Department of Astronomy, Cornell University, Ithaca, New York 14853, USA.}
\affiliation{Department of Phyiscs, Harvard University, Cambridge, Massachusetts 02138, USA}
\author{Nesar S. Ramachandra}
\affiliation{CPS Division, Argonne National Laboratory, 9700 South Cass Avenue, Lemont, Illinois 60439, USA.}
\affiliation{Kavli Institute for Cosmological Physics, University of Chicago, 5640 South Ellis Avenue, Chicago, Illinois 60637, USA}
\author{Agn\`es Fert\'e}
\affiliation{Jet Propulsion Laboratory, California Institute of Technology, 4800 Oak Grove Drive, Pasadena, CA 91109, USA}
\author{Jason Rhodes}
\affiliation{Jet Propulsion Laboratory, California Institute of Technology, 4800 Oak Grove Drive, Pasadena, CA 91109, USA}

\maketitle

\section{Introduction}

Weak gravitational lensing offers a unique test of gravity on cosmological scales. The precision of these cosmological tests of gravity will increase in the coming decade thanks to large amounts of precise data coming from Stage IV experiments including Euclid\footnote{\url{https://www.euclid-ec.org/}}~\cite{laureijs2011euclid}, the Nancy Grace Roman Space Telescope\footnote{\url{https://roman.gsfc.nasa.gov/}}~\cite{2015arXiv150303757S} and the Rubin Observatory\footnote{\url{https://www.lsst.org/}}. 
\par With the improved statistical precision of these surveys, extreme care must be taken to account for and remove systematic errors which could introduce biases in the parameter inference. Modeling biases are of particular concern as weak lensing is sensitive to scales down to $k \sim 10 \ h  {\rm Mpc} ^{-1}$ in the matter power spectrum, deep into the nonlinear regime~\cite{Taylor:2018nrc}.  While it is not the primary focus of this work, the impact of baryonic feedback on these scales is also highly uncertain~\cite{Huang:2018wpy}.
\par To perform parameter inference without loss of information, one must be able to model small scales over a large volume in cosmological parameter space. It is not possible to do this analytically~\cite{Osato:2018ldv}, but one can obtain the nonlinear power spectrum from an emulator~\cite{Heitmann:2013bra, Knabenhans:2018cng, Winther:2019mus} (or halo model~\cite{Mead:2015yca}) trained (calibrated) on a suite of $\mathcal{O} (100)$~\cite{Knabenhans:2018cng} high resolution N-body simulations, run over a large volume of cosmological parameter space. 
\par Simulations must be run with more than a trillion particles to meet the percent-level matter power spectrum accuracy requirements of upcoming surveys~\cite{Schneider:2015yka}. This makes a search for deviations from general relativity (GR) challenging as it is likely infeasible to run a suite of N-body simulations at this precision for every modified theory of gravity which we would like to test, particularly since we must also account for baryonic feedback. Thus, we must cut scales from the cosmic shear data vector. 
\par Care must be taken to retain useful information while removing poorly modeled small scales to avoid bias. The $k$-cut~\cite{Taylor:2018snp} cosmic shear estimator and its generalization to the combination of cosmic shear and galaxy clustering~\cite{Taylor:2020zcg}  ($k$-cut $3 \times 2$ point statistics~\cite{Taylor:2020imc}) were developed to meet these criteria.\footnote{These estimators achieve a similar aim to the nulling scheme developed in~\cite{Bernardeau:2013rda}.} The $k$-cut cosmic shear method works by applying the Bernardeau-Nishimichi-Taruya (BNT) transformation~\cite{Mandelbaum:2017dvy} to the tomographic weak lensing power spectra $C^{ij} (\ell)$. This reorganizes the information so that it is possible to relate angular scales, $\ell$, to physical scales, $k$, in the matter power spectrum. Then angular scale cuts correspond to  physical scale cuts, making it easier to remove small-scale data. 
\par In this paper we use this method to constrain Hu-Sawicki $f(R)$ gravity~\cite{Hu:2007nk} using the Hyper Suprime-Cam Year 1 cosmic shear data~\cite{Mandelbaum:2017dvy}. To generate the $f(R)$ matter power spectrum, $P(k,z)$, we use  the emulator developed in~\cite{Ramachandra:2020lue, valogiannis_georgios_2020_4058880}. As this emulator was trained on fast COLA (COmoving Lagrangian Acceleration) simulations~\cite{Tassev:2013pn, Valogiannis_2017}, it becomes inaccurate above $k = 1 \ h {\rm Mpc}^{-1}$. Therefore we take a $k$-cut at this scale removing the smaller-scale data while preserving useful information at larger scales.
\par The primary aim of this work is to  provide a road-map for future tests of modified gravity with Stage IV photometric surveys. We envisage a three step process:
\begin{itemize}
    \item {\bf Step 1:} For each theory of gravity, develop a model for the nonlinear power spectrum using an emulator trained on simulations as in this work, or using a halo model approach~\cite{Cataneo:2018cic,Giblin:2019iit,Cataneo:2019fjp,Bose:2020wch}.
    \item {\bf Step 2:} Determine the $k$-mode, $k^{\rm max}$, where the emulator fails i.e. the scale at which the accuracy of the power spectrum model fails to meet the requirements of the survey.
    \item {\bf Step 3:} Use $k$-cut cosmic shear to perform the inference removing sensitivity to scales with $k > k^{\rm max}$.
\end{itemize}

In this paper, we first review $k$-cut cosmic shear in Sec. \ref{theory}. We discuss the Subaru Hyper Suprime-Cam Year 1 Survey (hereafter HSCY1) data and covariance matrix in Sec. \ref{data}. Finally, in Sec. \ref{results}, we present the $k$-cut cosmic shear parameter constraints for both $f(R)$ gravity and $\Lambda$CDM before concluding in Sec.~\ref{sec:conclusion}. Unless explicitly stated otherwise, we take the Planck 2018 cosmology~\cite{Aghanim:2018eyx} as our baseline cosmology.

\section{Theory}\label{theory}

\subsection{Lensing power spectrum}

We use the lensing power spectrum, $C (\ell)$, to extract the two-point information of the shear field.
Under the Limber~\cite{LoVerde:2008re, Kitching:2016zkn}, spatially-flat universe~\cite{Taylor:2018qda}, flat-sky~\cite{Kitching:2016zkn}, reduced shear~\cite{Deshpande:2019sdl, Dodelson:2005ir} and Zel'dovich approximations ~\cite{Kitching:2016zkn} it is given by
\begin{equation}
C_{GG}^{ij}(\ell) = \int_0^{\chi_H}d\chi \frac{q^i(\chi)q^j(\chi)}{\chi^2} P\left(\frac{\ell+1/2}{\chi},\chi\right).
\label{eq:CGG}
\end{equation}
Here $\chi$ is the radial comoving distance, $P$ is the matter power spectrum, $\chi_H$ is the distance to the horizon and $q(\chi)$ is the lensing efficiency kernel, defined as:
\begin{equation}
q^i(\chi) = \frac{3}{2}\Omega_m \left(\frac{H_0}{c}\right)^2 \frac{\chi}{a(\chi)} \int_{\chi}^{\chi_H} d\chi' n^i(\chi') \frac{\chi'-\chi}{\chi'}.
\end{equation}
\par In this expression, $H_0$ is the Hubble constant, $\Omega_m$ is the fractional matter density parameter today, $c$ is the speed of light, $a$ is the scale factor and $n^i(\chi')$ is the probability distribution of the effective number density of galaxies as a function of comoving distance. We also account for intrinsic alignments due to galaxy alignments with the local tidal field. The total angular power spectrum is given by
\begin{equation}
    C^{ij} (\ell) = C_{GG}^{ij}(\ell) + C_{IG}^{ij}(\ell) + C_{II}^{ij}(\ell),
\end{equation}
where $ C_{IG}^{ij}(\ell)$ accounts for correlations between tidally aligned foreground galaxies with gravitationally sheared background galaxies and $ C_{II}^{ij}(\ell)$ accounts for the autocorrelation between local tidally aligned galaxies. These are given by 
\begin{equation}
    C_{IG}^{ij}(\ell) = \int_{0}^{\chi_H} d\chi \frac{n^i(\chi)q^j(\chi)}{\chi^2}P_{IG}\left( \frac{\ell+1/2}{\chi}, \chi \right),
\end{equation}
and
\begin{equation}
    C_{II}^{ij}(\ell) = \int_{0}^{\chi_H} d\chi \frac{n^i(\chi)n^j(\chi)}{\chi^2}P_{II}\left( \frac{\ell+1/2}{\chi}, \chi \right).
\end{equation}
\par We use the extended nonlinear alignment model (eNLA)~\cite{joachimi2011constraints} used in the HSCY1 analysis ~\cite{Hikage:2018qbn} and Dark Energy Survey (DES) Year 1 analysis~\cite{Troxel:2017xyo}. This is similar to the standard nonlinear alignment model (NLA)~\cite{Bridle:2007ft}, but the amplitude of the intrinsic alignments, $A_{\rm IA} (z)$ is allowed to  vary with redshift following
\begin{equation} \label{eqn:Fz} 
A_{\rm IA}(z) = - A_{\rm IA} \left[ \frac{1 + z}{1 + z_0}  \right] ^ \alpha,
\end{equation}
where $A_{\rm IA}$ and $\alpha$ are free nuisance parameters and we fix $z_0 = 0.62$~\cite{Hikage_2019}. 
\par In all that follows we use {\tt CosmoSIS}~\cite{Zuntz:2014csq} to compute the lensing spectrum using {\tt CAMB}~\cite{Lewis:1999bs} to generate the linear power spectrum and {\tt Halofit}~\cite{Takahashi:2012em} to generate the $\Lambda$CDM nonlinear power spectrum. We describe how the nonlinear $f(R)$ power spectrum is generated in Sec. ~\ref{sec:emulator}.

\subsection{The BNT transform and $k$-cut cosmic shear} \label{BNT}

\begin{figure} [hbt!]
    \centering
    \includegraphics[width = \columnwidth]{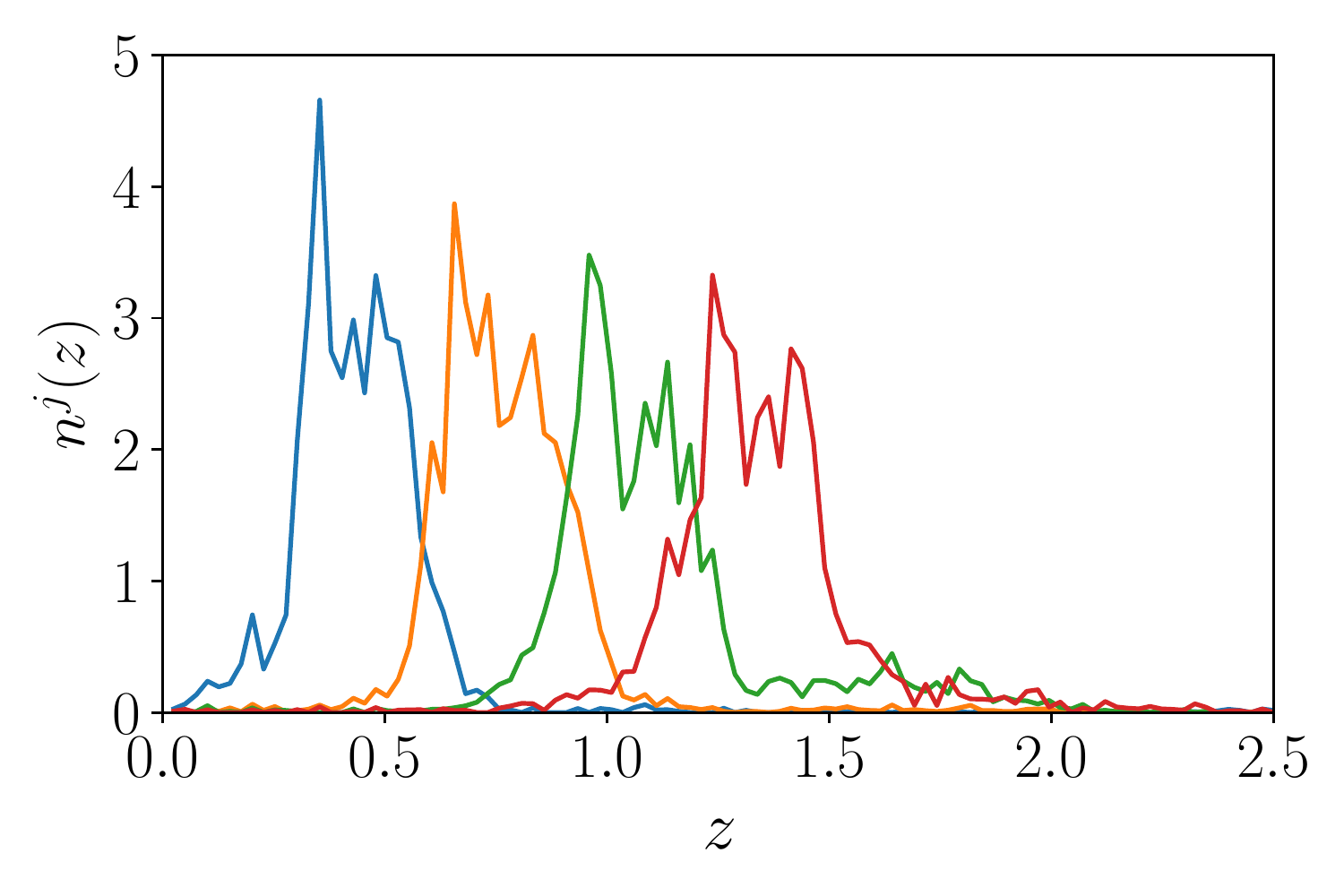}
    \includegraphics[width=\columnwidth]{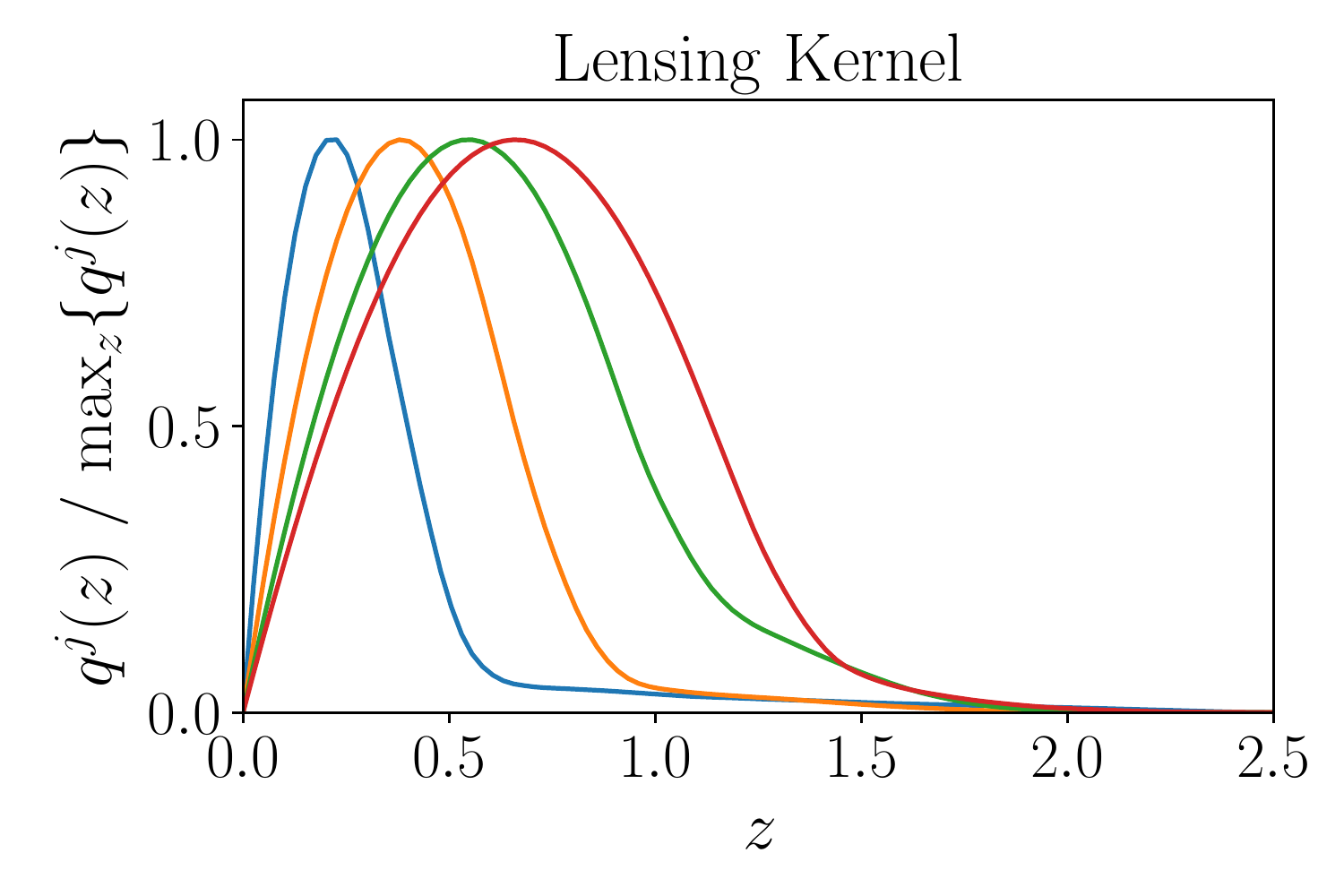}
    \includegraphics[width=\columnwidth]{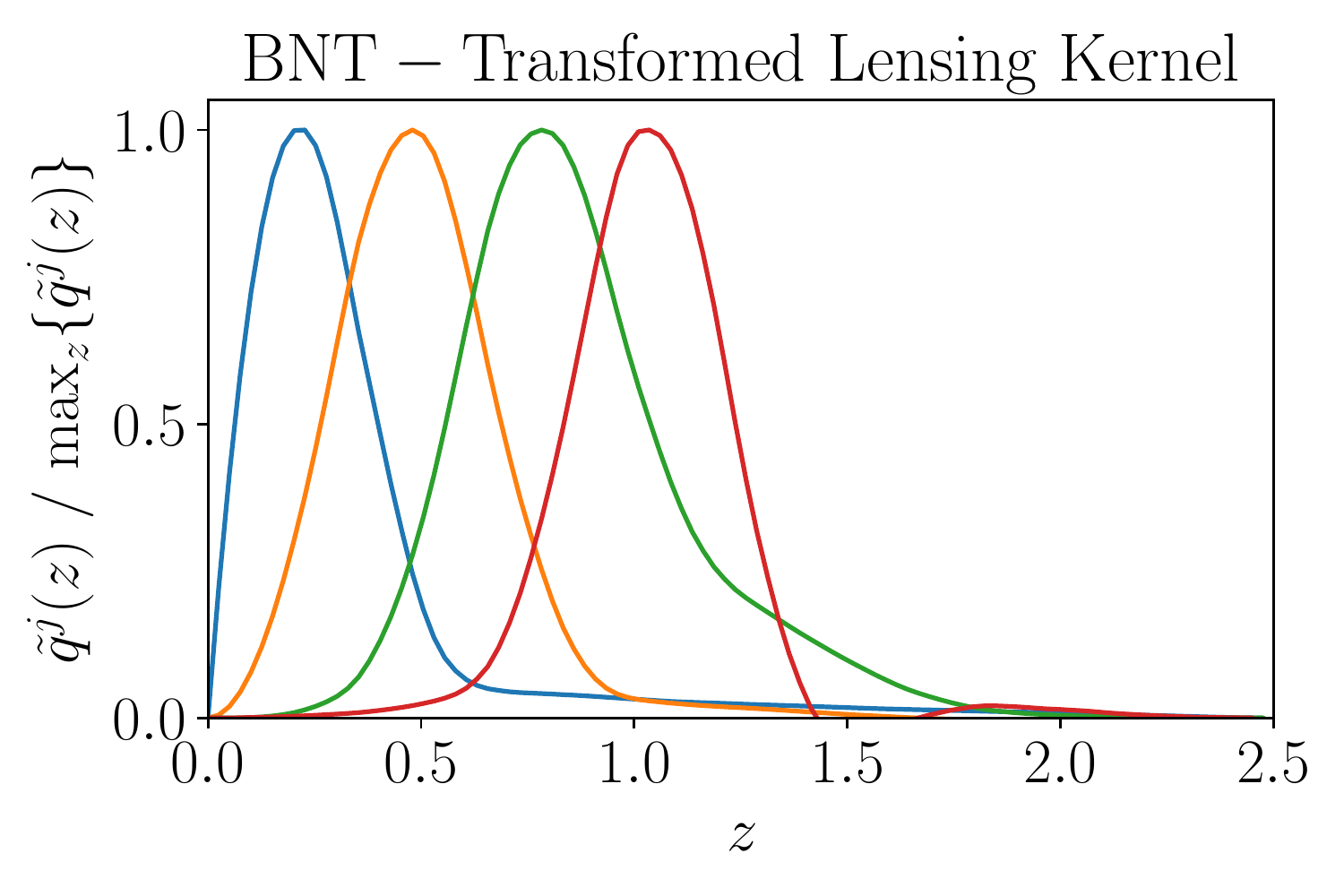}

    \caption{{\bf Top:} radial distribution function for the four HSCY1 tomographic bins. {\bf Middle:} the corresponding lensing efficiency kernels, $q^j(z)$, normalized against their maximum values. {\bf Bottom:} the BNT-transformed lensing kernels, $\widetilde q^j(z)$. These are narrower in $z$ allowing us to related physical and angular scales through the Limber (small-angle) approximation $\ell \sim kr$. Cuts in $\ell$ then correspond to physical scale cuts.}
    \label{lensing_cuts}
\end{figure}

\begin{figure*} [hbt!]
    \centering
    \includegraphics[width=\linewidth]{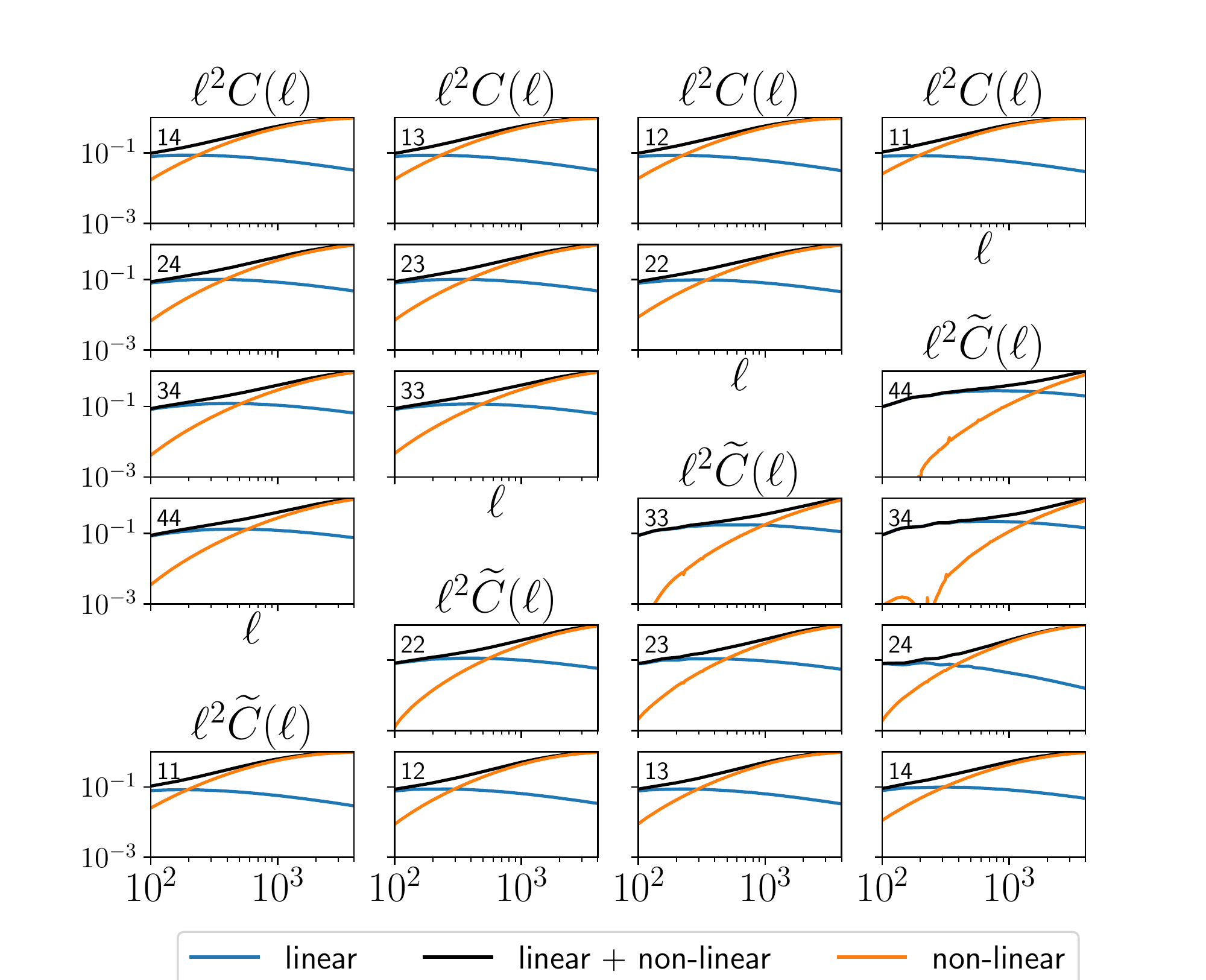}
    \caption{{\bf Top left:} tomographic theoretical angular power spectra, $C(\ell)$ for the best-fit cosmology of HSCY1. The tomographic bin combination is indicated in the top left of each subplot. Black lines are used for the total angular power spectra, while blue and orange lines are used to indicate the linear and nonlinear contributions from the nonlinear matter power spectrum respectively. {\bf Bottom right:} BNT-transformed angular power spectra, $\widetilde C (\ell)$. Due to the sorting of physical scales, the nonlinear contributions become important at higher-$\ell$ modes than for standard tomographic angular power spectra. This implies we can take a cut at a higher-$\ell$ mode and extract more information while remaining robust to modeling uncertainties at small physical scales.}
    \label{kcut}
\end{figure*}
Weakly lensed galaxies probe structure along the entire line-of-sight. Lensing structures of the same physical scale at different redshifts subtends different angles on the sky, breaking the correspondence between angular and physical scales. Hence, taking an angular scale cut does not correspond to taking a physical scale cut. 
\par In order to find a more precise correspondence between physical and angular scales, in the $k$-cut method~\cite{Taylor:2018snp,Taylor:2020imc} we make use of the Bernardeau-Nishimichi-Taruya (BNT) transform~\cite{Bernardeau:2013rda}. Specifically we take a linear combination of lensing kernels, $q^i$, so that
\begin{equation}
\tilde q^i(z) = M^{ij} q^j(z),
\end{equation}
where we sum over repeated indices, $M^{ij}$ is the BNT matrix\footnote{The BNT matrix is implicitly a function of cosmology, but we compute it once at the fiducial cosmology and leave it fixed for the remainder of this work. It was found in~\cite{Taylor:2020zcg} that the choice of fiducial cosmology has a negligible impact.} and $\tilde q^i (z)$ are the BNT lensing kernels. 
\par The BNT matrix is found by first setting $M^{ii} = 1$ for all $i$ and $M^{ij} = 0$ for $i<j$, the remaining BNT matrix elements are found by solving the system
\begin{equation}
\begin{aligned}
\sum_{j=i-2}^{i}M^{ij} &= 0\\
\sum_{j=i-2}^{i}M^{ij}B^j &= 0,
\end{aligned}
\end{equation}
where
\begin{equation}
    B^j = \int_0^{z_{\rm max}} {\rm d}z' \frac{n^j (z')}{\chi(z')}
\end{equation}
and $z_{\rm max}$ is the maximum redshift of the survey. We use the publicly available code at: \hyperlink{https://github.com/pltaylor16/x-cut}{https://github.com/pltaylor16/x-cut} to compute the BNT matrix.
\par The top row of Fig.~\ref{lensing_cuts} shows the four photometric redshift bins used in the HSCY1 cosmic shear analysis~\cite{Hikage_2019}. The corresponding lensing kernels are plotted in the second row. The kernels are broad in $z$, which as we have argued makes it difficult to relate physical and angular scales. Meanwhile the BNT kernels are shown in the bottom row of Fig.~\ref{lensing_cuts}. These are much narrower in $z$, which will allow us to relate physical and angular scales at the two-point level.
 \par Since the lensing power spectrum depends on two sets of kernels labeled by tomographic bin numbers $i$ and $j$, the correct transform at the two-point level is 
\begin{equation}
    \widetilde C^{ij}(\ell)= M^{ik} C(\ell)^{kl} \left( M^T \right)^{lj},
\end{equation}
where we sum over repeated indices. We refer to $\widetilde C^{ij}(\ell)$ as the $k$-cut cosmic shear power spectrum. 
\par The power spectra (top right) and BNT-transformed power spectra (bottom right) are shown in Fig.~\ref{kcut}. The linear and nonlinear contributions are shown in blue and orange respectively and the combination of the two is shown in black. We expect a bigger change between the BNT-transformed power spectrum and the standard power spectrum for higher redshift correlation, which we indeed find. Crucially the linear and non-linear power spectra contributions are comparable at higher-$\ell$ in the BNT-transformed case. This implies that we can take scale cuts at higher $\ell$-values in the BNT-transformed case, while still removing sensitivity to poorly modeled nonlinear scales.
\par Formally, since each kernel is narrow in $z$, we can define a ``typical distance,'' $\widetilde \chi^i$, which is given by comoving distance at the redshift where kernel $\widetilde q^i(z)$ obtains its maximum value\footnote{One could alternatively take $\widetilde \chi^i$ as the co-moving distance at the average redshift over the kernel.}, assuming a fiducial cosmology.
Then if we want to remove sensitivity to scales smaller than some $k$-mode in the matter power spectrum, we can use the Limber relation and cut all $\ell > k \widetilde \chi ^{ij}$, where $\widetilde \chi ^{ij} = \min (\widetilde \chi ^i, \widetilde \chi ^j)$. This procedure is referred to as $k$-cut cosmic shear. 
\par The effectiveness of the method at removing sensitivity to small scales comes with several caveats. The BNT-transformed kernels, $\widetilde q^j(z)$, shown in Fig.~\ref{lensing_cuts} still have some width so the $\ell$ -- $k$ correspondence is not exact. This will become less problematic in future surveys as the number of tomographic bins is increased and photometric redshift errors improve. Since the BNT transform depends on $n^j(z)$ the effectiveness of the transform also relies on the accuracy of the photometric redshift estimates.\footnote{Formally the BNT transform also depends on the background cosmology through the mapping from $z$ to $\chi$, but in practice it was shown in~\cite{Taylor:2020zcg} that the transform is extremely insensitive to the choice of fiducial cosmology.} In light of these issues, we perform a test in Sec.~\ref{sec:verification} to ensure that $k$-cut method cuts sensitivity to small scales as intended.

\subsection{$k$-cut cosmic shear likelihood and covariance}\label{cov}
We develop our own module to evaluate the $k$-cut cosmic shear likelihood in {\tt CosmoSIS}~\cite{Zuntz_2015}, a modular open-source package for likelihood-based cosmological parameter inference.
\par We assume that the $k$-cut cosmic shear likelihood is Gaussian. This will need to be verified in a follow-up study following the method in~\cite{Taylor:2019mgj} or~\cite{Upham:2020klf}. The data and theory vectors are found by BNT transforming the theory and data vectors of the standard angular power spectra, $C^{ij} (\ell)$. 
\par The covariance matrix of the BNT transformed data vector, $ \widehat C_{\rm BNT}^{ab,cd}(\ell, \ell')$, can be computed from an estimate of the covariance of the tomographic angular power spectra. The elements are given by
\begin{equation} \label{eqn:cov}
    \widehat C_{\rm BNT}^{ab,cd}(\ell, \ell') = M^{a e} M^{b f} M^{c g} M^{d h}  \widehat C^{ef,gh}(\ell, \ell'),
\end{equation}
where $\widehat C^{ef,gh}(\ell, \ell')$ denotes the covariance between $C^{ef}(\ell)$ and $C^{gh}(\ell')$. We list analogous expressions for the covariance elements in the $k$-cut $3 \times 2$ point formalism~\cite{Taylor:2020imc} in the Appendix~\ref{sec:cov2}. One can also use the likelihood sampling method to compute the covariance of the BNT-transformed data vector (see~\cite{taylor2020xcut} for more details).
\par In this paper we used the covariance matrix, $\widehat C^{ef,gh}(\ell, \ell')$, computed in~\cite{Hikage_2019} at the best-fit cosmology therein. This covariance includes all Gaussian, non-Gaussian and super-sample covariance terms. We refer the reader to the Appendix of~\cite{Hikage:2018qbn} for more details.

\par
\begin{figure} [hbt!]
    \centering
    \includegraphics[width=\columnwidth]{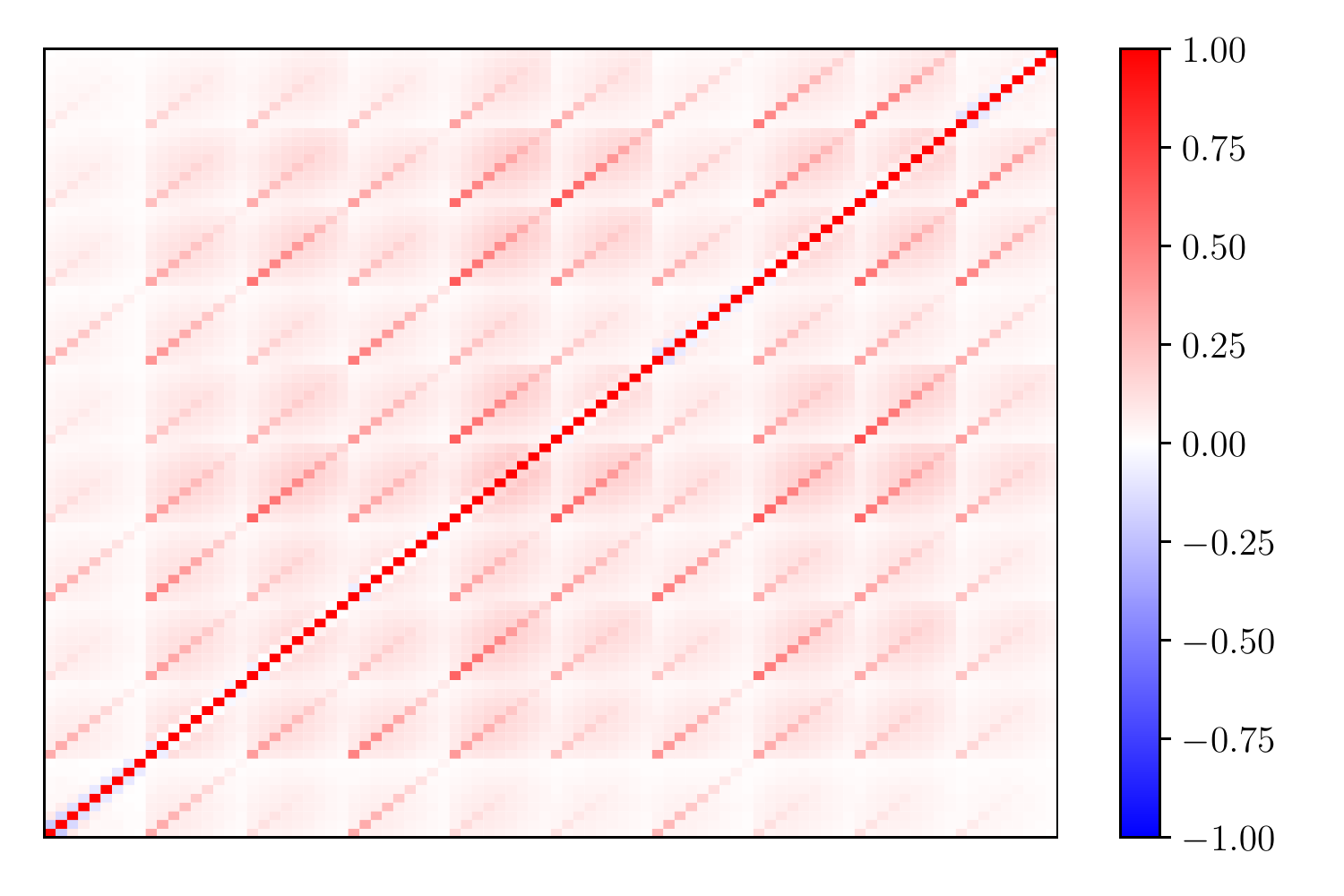}
    \includegraphics[width=\columnwidth]{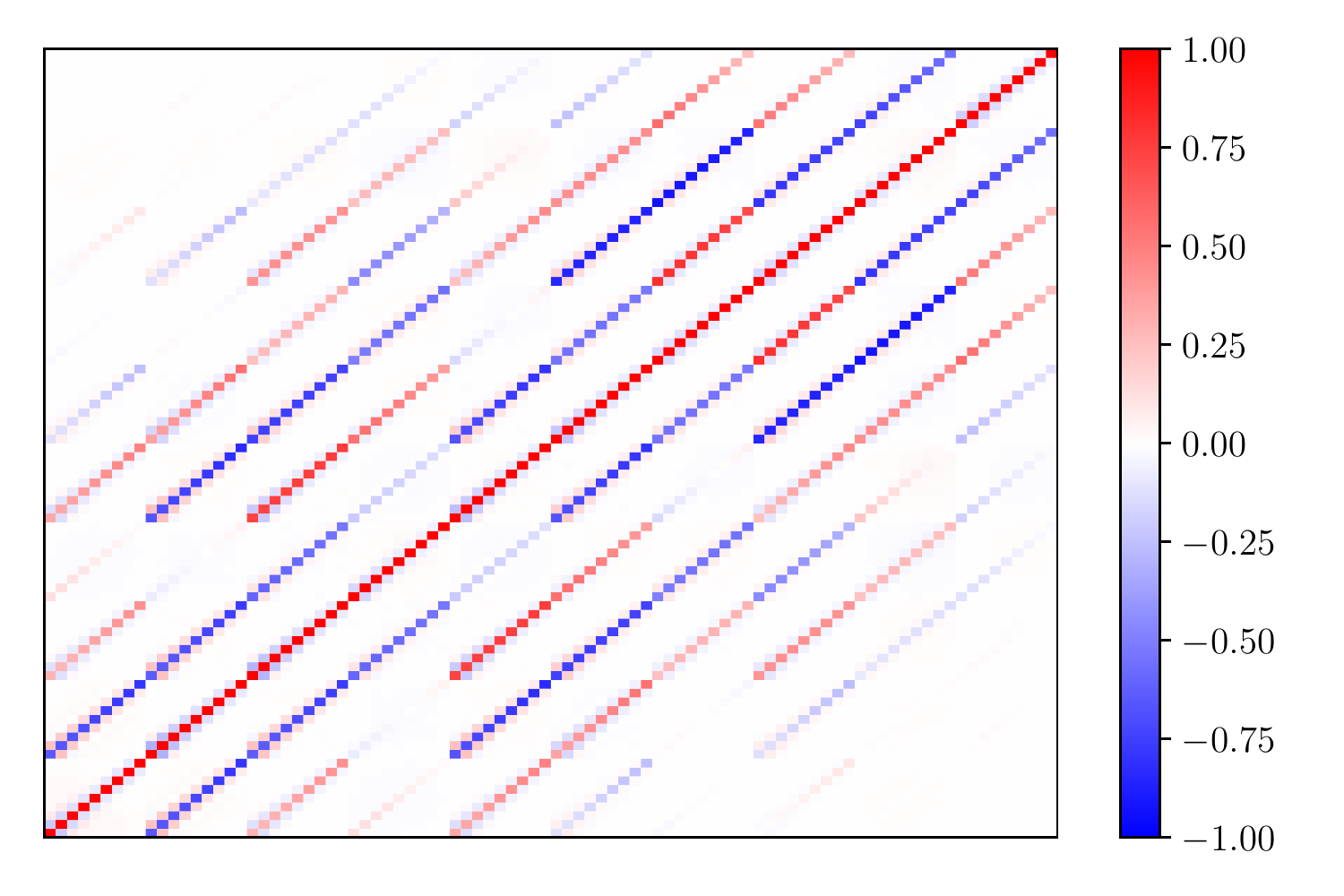}
    \caption{{\bf Top:} the angular power spectra correlation matrix from~\cite{Hikage_2019}, computed at the best fit cosmology from the HSCY1 analysis. The matrices are ordered into block matrices with tomographic bin ordering $\{i,j \} = 11, 12, 13, 14, 21, 22, 23, 24, 33, 34, 44$ and increasing $\ell$ inside each block. {\bf Bottom:} the $k$-cut cosmic shear correlation matrix. As expected the $k$-cut covariance is much sparser as the BNT transform reduces the redshift overlap between the lensing efficiency kernels (see Fig.~\ref{lensing_cuts}) weakening the covariance between tomographic bins.}
    \label{BNT_cov}
\end{figure}

 The HSCY1 and $k$-cut cosmic shear correlation matrices\footnote{For a covariance matrix, $C$, the correlation matrix is  $C^{ij} = C^{ij} / \sqrt{C^{ii}}\sqrt{C^{jj}}$} are shown in Fig. \ref{BNT_cov}. The matrices are ordered into block matrices with tomographic bins ordering $\{i,j \} = 11, 12, 13, 14, 21, 22, 23, 24, 33, 34, 44$ and increasing $\ell$ inside each block. The $k$-cut covariance is much sparser. This is expected as by construction the BNT transform reduces the redshift overlap between the lensing efficiency kernels (see Fig.~\ref{lensing_cuts}) weakening the covariance between different tomographic bins.

\subsection{$f(R)$ gravity}
In $f(R)$ gravity, the Einstein-Hilbert action $S$ is modified by including an 
extra function of the Ricci curvature, $f(R)$ \cite{DeFelice:2010aj} so that,
\begin{equation}
S = \int{d^4x\sqrt{-g}\left[\frac{R+f(R)}{16 \pi G}+ \mathcal{L}_M\right]},
\end{equation}
where $\mathcal{L}_M$ gives the Lagrangian of matter in the Universe, $G$ the gravitational constant and the speed of light, $c$, is set to unity. In principle the extra degrees of freedom in this model can drive cosmic acceleration.
\par In this paper we consider the Hu-Sawicki parametrization~\cite{Hu:2007nk}, which to date has passed all observational tests in the GR limit~\cite{Burrage:2017qrf} and is well-studied in the literature, making it an ideal choice for the  first application of the $k$-cut framework to a test of modified gravity. In this parametrization
\begin{equation}
    f(R) = -m^2 \frac{c_1 (R/m^2)^n}{c_2 (R/m^2)^n + 1},
\end{equation}
where $m = H_0 \sqrt{\Omega_m}$ and $n$, $c_1$ and $c_2$ are free parameters. Matching the background expansion to that of a $\Lambda$CDM universe imposes as additional constraint on the derivative of $f(R)$ with respect to background Ricci scalar, $\frac{\partial f(R)}{\partial R}$, evaluated today
\begin{equation}
f_{R_0}  = -n \frac{c_1}{c_2 ^ 2} \left( \frac{\Omega_m}{3 \left( \Omega_m + \Omega_\Lambda\right)}\right).
\end{equation}
The Hu-Sawicki model is typically parameterized in terms of two free parameters: $\left| f_{R_0} \right|$ and $n$,\footnote{For convenience we will write $\left| f_{R_0} \right|$ as $ f_{R_0}$ for the remainder of this work.} and the theory reduces to GR in the limit $f_{R_0}  \rightarrow 0$.

\subsection{$f(R)$ power spectrum emulator} \label{sec:emulator}
\par We use the $f(R)$ Hu-Sawicki modified gravity Gaussian process emulator developed in~\cite{Ramachandra:2020lue} and available at \url{https://github.com/LSSTDESC/mgemu} to compute the nonlinear $f(R)$ power spectrum. This emulator was trained inside a five-dimensional parameter space using a suite of $f(R)$ comoving Lagrangian acceleration (COLA) simulations~\cite{Valogiannis_2017}. The parameters and their prior ranges are listed in Table~\ref{tab:fr_prior}. The Hubble parameter is fixed as $h_0 = 0.67$ as assumed in~\cite{Ramachandra:2020lue}. Note in particular that we restrict $\log (f_{R_0})$ to a flat prior with $f_{R_0} \in [10^{-4}, 10^{-8}]$ as this is the range in which the $f(R)$ emulator is accurate~\cite{Ramachandra:2020lue}.

\begin{table} 
\begin{tabular}{ || c | c ||}
\hline
     Parameter & Prior  \\
     \hline
     \hline
     $\log(f_{R_0})$ & flat[-8, -4] \\
     $n$ & flat[0, 4] \\
     $\Omega_mh^2$ & flat[0.12, 0.15] \\
     $n_s$ & flat[0.85, 0.1] \\
     $\sigma_8$ & flat [0.7, 0.9] \\
\hline     
\end{tabular}
\caption{ The cosmological parameter space where the $f(R)$ emulator is valid.  In the $f(R)$ likelihood analysis, these ranges are used as the cosmology prior while the priors on intrinsic alignments and nuisance parameters are given in the bottom two segments of Table.~\ref{tab:2}.}
\label{tab:fr_prior}
\end{table}

\par COLA uses a combination of second order Lagrangian perturbation theory (2LPT) and an N-body component to retain most of the accuracy of full N-body simulations at a fraction of the computational cost. The emulator is accurate at the $5 \%$ level up to $k = 1 \ h {\rm Mpc} ^{-1}$, achieving $1 \%$ accuracy for $f_{R_0} < 10^{-5}$. We refer the reader to~\cite{Ramachandra:2020lue} for more details. 
\par The emulator computes the ratio of the $f(R)$ and $\Lambda$CDM nonlinear power spectra i.e. $P_{f(R)}(k,z)/ P_{\Lambda\rm{CDM}} (k,z)$. We use {\tt Halofit}~\cite{Takahashi:2012em} to compute $P_{\Lambda\rm{CDM}} (k,z)$ and we have developed our own {\tt CosmoSIS} module to call the emulator and compute $P_{f(R)}(k,z)$. 
\par Although baryonic feedback represents a subdominant effect on scales larger than $1 \ h {\rm Mpc}^{-1}$~\cite{Huang:2018wpy}, it is noted in Appendix A of~\cite{Hikage_2019} that inaccuracies of the {\tt Halofit} dark matter power spectrum model can result in $1 \sigma$ biases on $\sigma_8$ for a cut at $\ell = 2000$. We expect this to serve as an upper bound on the biases in this $k$-cut study since~\cite{Mead:2015yca} found that the {\tt Halofit} model is accurate to within $\sim 4 \%$ below $1 \ h {\rm Mpc}^{-1}$ and typically accurate to $\sim 2 \%$ below $1 \ h {\rm Mpc}^{-1}$ over most of this $k$-range. As we will show in Sec.~\ref{sec:lcdm} this is smaller than our measurement error on $S_8$. Nevertheless future studies can sidestep this issue by using a more accurate emulator such as~\cite{Euclid:2018mlb} or~\cite{Angulo:2020vky}.

\section{Data}\label{data}

\subsection{Hyper Suprime-Cam year 1 data}

We use data from the first year of the Hyper Suprime-Cam (HSC) survey~\cite{Mandelbaum:2017dvy}. The data covers $136.9 \ {\rm deg}^2$ over 6 disjoint fields with an effective source number density of $17$ galaxies per ${\rm arcmin} ^2$~\cite{Hikage_2019}. We refer the reader to~\cite{Mandelbaum:2017dvy} for a detailed description of the shear estimation. 
\par Galaxies were binned into 4 redshift bins inside the range $0.3 < z < 1.5$ with a magnitude cut of $i < 24.5$~\cite{Hikage_2019} and redshifts were estimated from the COSMOS 30-band photo-z catalog (see~\cite{Hikage:2018qbn,Gruen:2016jmj,Ilbert:2008hz,tanaka2018photometric,Masters:2015asa} for more details). The resulting redshift probability density functions, $n^j(z)$, for each bin are plotted in Fig.~\ref{lensing_cuts}. Power spectra were estimated in 15 logarithmically spaced bandpowers in the range $30<\ell<6500$ using the pseudo-$C_\ell$ pipeline described in Sec.~\ref{BNT} . We cut all $\ell < 300$ as in~\cite{Hikage_2019} and take additional $\ell$-cuts after applying the BNT transform following the $k$-cut cosmic shear procedure. In this paper we remove sensitivity to all scales corresponding to $k$ larger than $ k = 1 \ h {\rm Mpc} ^{-1}$ which corresponds to cutting bandpowers with effective $\ell > 1205, \ 1814, \ 2329$ and $2781$ for the four tomographic bins respectively. In contrast the HSCY1 analysis took a global multipole cut at $\ell = 2000$~\cite{Hikage_2019}.

\subsection{Residual systematics}
As in~\cite{Hikage:2018qbn}, we consider two sources of residual bias: multiplicative shear bias and linear photometric redshift error.
\par The theoretical angular power spectra are rescaled by a multiplicative factor
\begin{equation}
    C ^{ij} (\ell) \rightarrow \delta m^i \delta m^j C^{ij} (\ell)
\end{equation}
where
\begin{equation}
\delta m^ i = (1+\Delta m^i)(1+m^i_{\rm sel} + m^i_\mathcal{R}).
\end{equation}
Here $m^i_{\rm sel}$ and $m^i_\mathcal{R}$ are fixed parameters which account for the selection bias and a residual bias in the shear response in tomographic bin $i$ (see~\cite{Hikage:2018qbn} Sec. 5.7 for more details), while $\Delta m^i$ is a residual multiplicative left as a free parameter in this analysis. We use the same values and prior ranges as in~\cite{Hikage:2018qbn}\footnote{Unlike in~\cite{Hikage:2018qbn} each tomographic bin is assigned its own residual multiplicative bias as there is no \textit{a priori} reason to assume that they should be the same for each bin}, but they are outlined in Table~\ref{tab:2} for convenience.
\par We also allow for a linear bias in the redshift probability distribution function
\begin{equation}
    n^i(z) \rightarrow n^i(z + \Delta z ^ i).
\end{equation}
We use the same prior ranges for $\Delta z^i$ as in the fiducial HSCY1~\cite{Hikage:2018qbn} analysis. They are displayed in Table~\ref{tab:2}.
\par We do not marginalize over any free parameters to account for point spread function residuals, as this was found to have negligible impact in~\cite{Hikage_2019}.

\section{Results} \label{results}

\subsection{Verification of $k$-cut method } \label{sec:verification}

In this section we perform a test to ensure the $k$-cut method removes sensitivity to the desired scales. We fix the cosmological parameters choosing a Planck 2018 cosmology and constrain the amplitude, $\mathcal{A}_P$, of the power spectrum above the target $k_{\rm cut} = 1 \ h {\rm Mpc} ^{-1}$ using: 1) a $k$-cut analysis and 2) a standard $C (\ell)$ analysis. In both cases the largest bandpower has an effective $\ell = 3080$. The amplitude, $\mathcal{A}_P$, is taken relative to the Planck 2018 prediction so that $\mathcal{A}_P = 1$ corresponds to the matter power spectrum for a Planck cosmology. Since the $k$-cut data vector should be insensitive to modes $k_{\rm cut} > 1 \ h {\rm Mpc} ^{-1}$ by construction, if the errors on $\mathcal{A}_P$ are significantly larger in the $k$-cut case, this implies that small scales are nulled as intended. 
\par Formally we rescale the power spectrum
\begin{equation}
    P(k,z) \rightarrow \big[ 1 +  (\mathcal{A}_P - 1) \Theta(k_{\rm cut}) \big] P(k,z)
\end{equation}
where 
\begin{equation}
  \Theta(k_{\rm cut}) =\left\{
  \begin{array}{@{}ll@{}}
    0, & \text{if} \ k < k_{\rm cut} \\
    1, & \text{if} \ k \geq k_{\rm cut}\\
  \end{array}\right.
\end{equation}
and use the {\tt Emcee} sampler~\cite{foreman2013emcee} in {\tt CosmoSIS} to constrain $\mathcal{A}_P$.
\par
\begin{figure} [hbt!]
    \centering
        \includegraphics[width=\columnwidth]{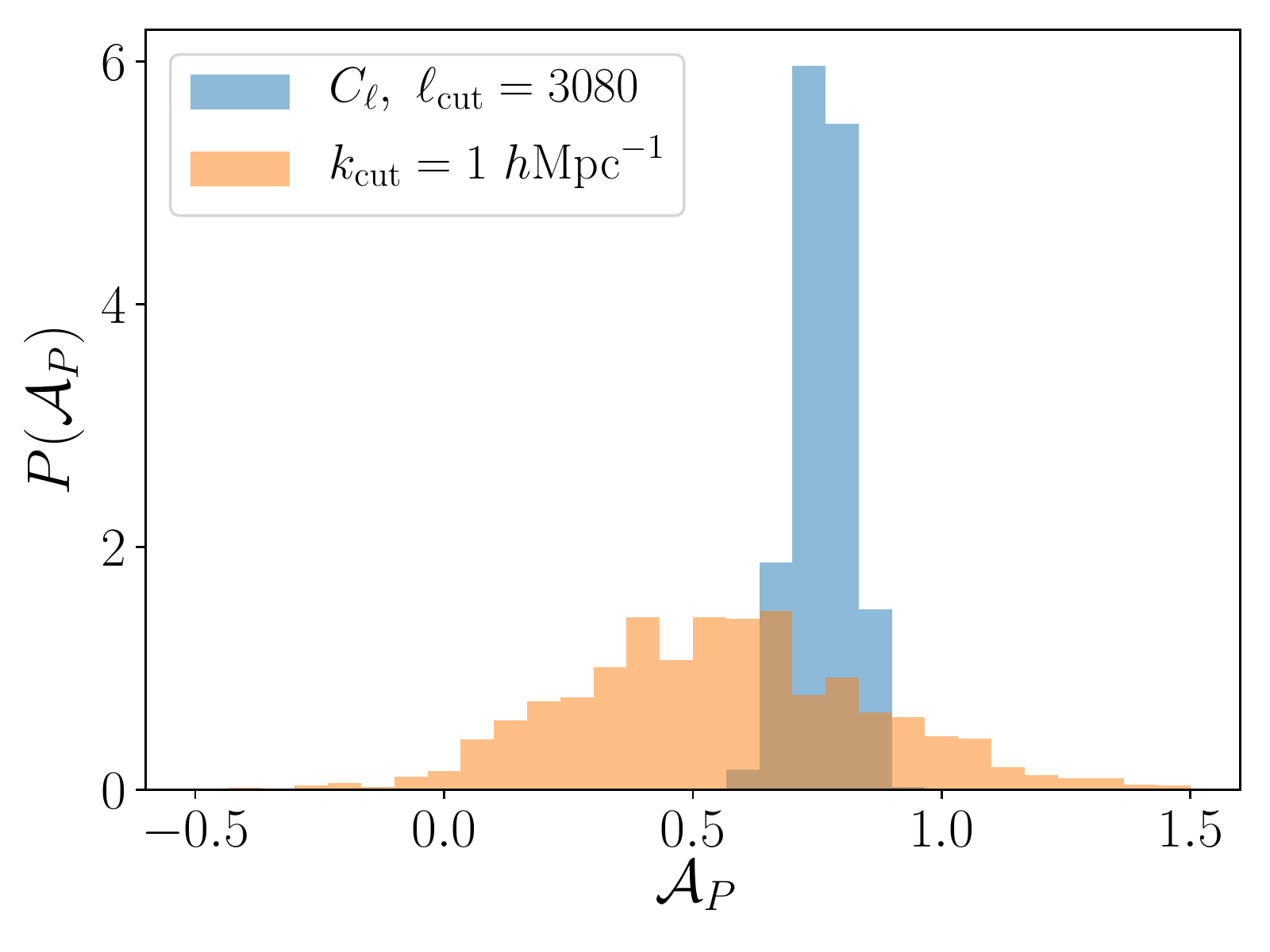}
    \caption{Constraints on the amplitude of the power spectrum above $k = 1 \ h {\rm Mpc}^{-1}$ with all other parameters fixed. {\bf Blue:} standard power spectrum analysis. {\bf Orange:} $k$-cut cosmic shear analysis. The error is 5.5 times larger in the $k$-cut case demonstrating that small uncertain scales are nulled as intended.}
    \label{fig:test cut}
\end{figure}

 The results of this analysis are shown in Fig.~\ref{fig:test cut}. The error on the amplitude, $\sigma (\mathcal{A}_P)$ increases from $0.06$ to $0.31$ or a factor of $5.5$ validating that sensitivity to small scales is significantly down-weighted. We recommend this validation step is performed in all future $k$-cut analyses. 
\par A similar rescaling of the power spectrum was used to search for deviations from $\Lambda$CDM in~\cite{Taylor:2018mqm} using The Canada-France-Hawaii Telescope Lensing Survey (CFHTLenS) data. We find no evidence that our constraints on  $\mathcal{A}_p$ are inconsistent with $\Lambda$CDM. Although we find the majority of the posterior lies below $\mathcal{A}_p < 1$ in both the $C (\ell)$ and $k$-cut analysis, this is the expected behavior for two reasons. First, we did not include baryonic physics which is known to suppress the power spectrum above $k = 1 \ h {\rm Mpc}$. Second, we use the Planck 2018 cosmology. Weak lensing constraints are known to favor lower values for $\sigma_8$, i.e. the amplitude of the power spectrum. 

 \begin{figure} [hbt!]
    \centering
    \includegraphics[width=\columnwidth]{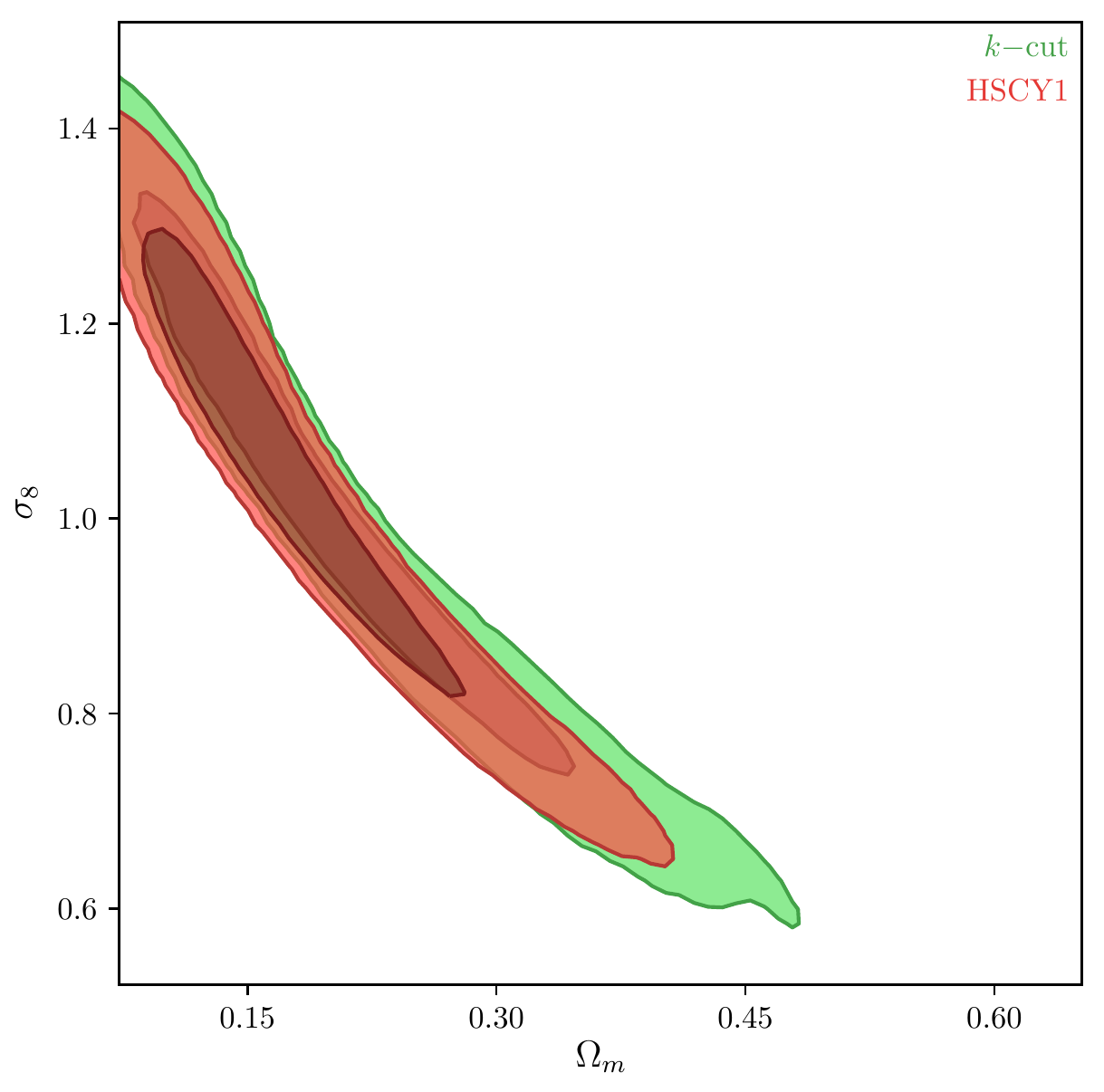}
    \includegraphics[width=\columnwidth]{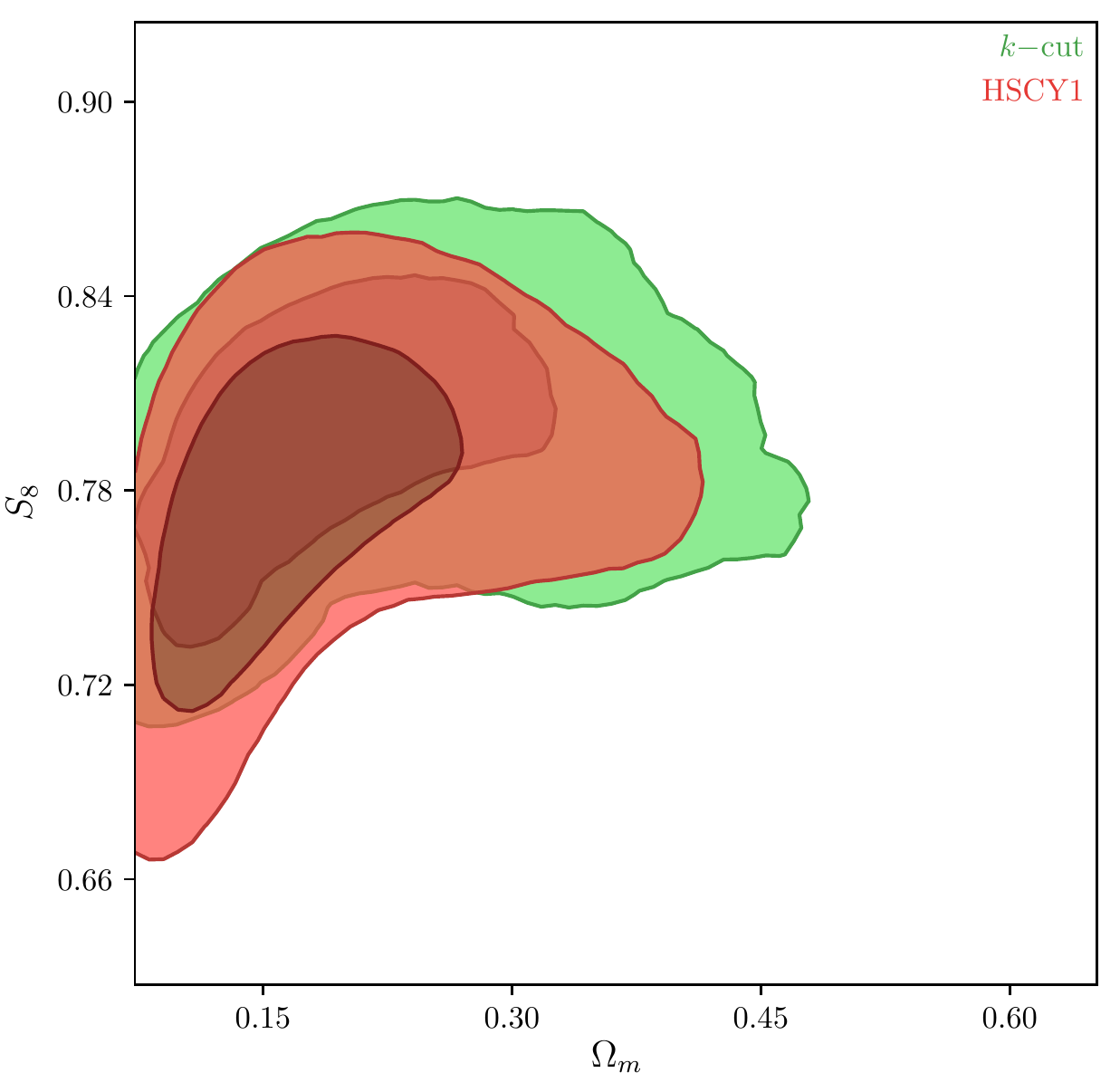}
    \caption{A comparison of the $k$-cut $\Lambda$CDM constraints in this analysis (green) with the HSCY1 fiducial analysis (red). Despite a conservative choice of scale cut at $k = 1 \ h {\rm Mpc} ^ {-1}$, the $k$-cut method extracts the majority of the information. The symmetric error on $S_8$ is $3 \%$ larger in the $k$-cut constraints compared to the fiducial analysis.}
    \label{fig:s8 compare}
\end{figure}

\subsection{$\Lambda$CDM comparison with HSCY1 analysis} \label{sec:lcdm}
Using our {\tt CosmoSIS} pipeline and the {\tt Multinest} sampler~\cite{Feroz_2009}, we perform the first $k$-cut likelihood analysis using real data with a target $k_{\rm cut} = 1 \ h {\rm Mpc} ^{-1}$. This scale cut was chosen as it is the point where the $f(R)$ power spectrum becomes inaccurate~\cite{Ramachandra:2020lue} to maintain consistency with our $f(R)$ results. Conveniently this choice also removes sensitivity to scales where baryonic corrections to the nonlinear matter power spectrum~\cite{Huang:2018wpy, Taylor:2020zcg} become important. Thus we use the dark matter-only power spectrum without marginalizing over any additional nuisance parameters to account for baryonic feedback. 
\par
\begin{figure*} [hbt!]
    \centering
    \includegraphics[width=\textwidth]{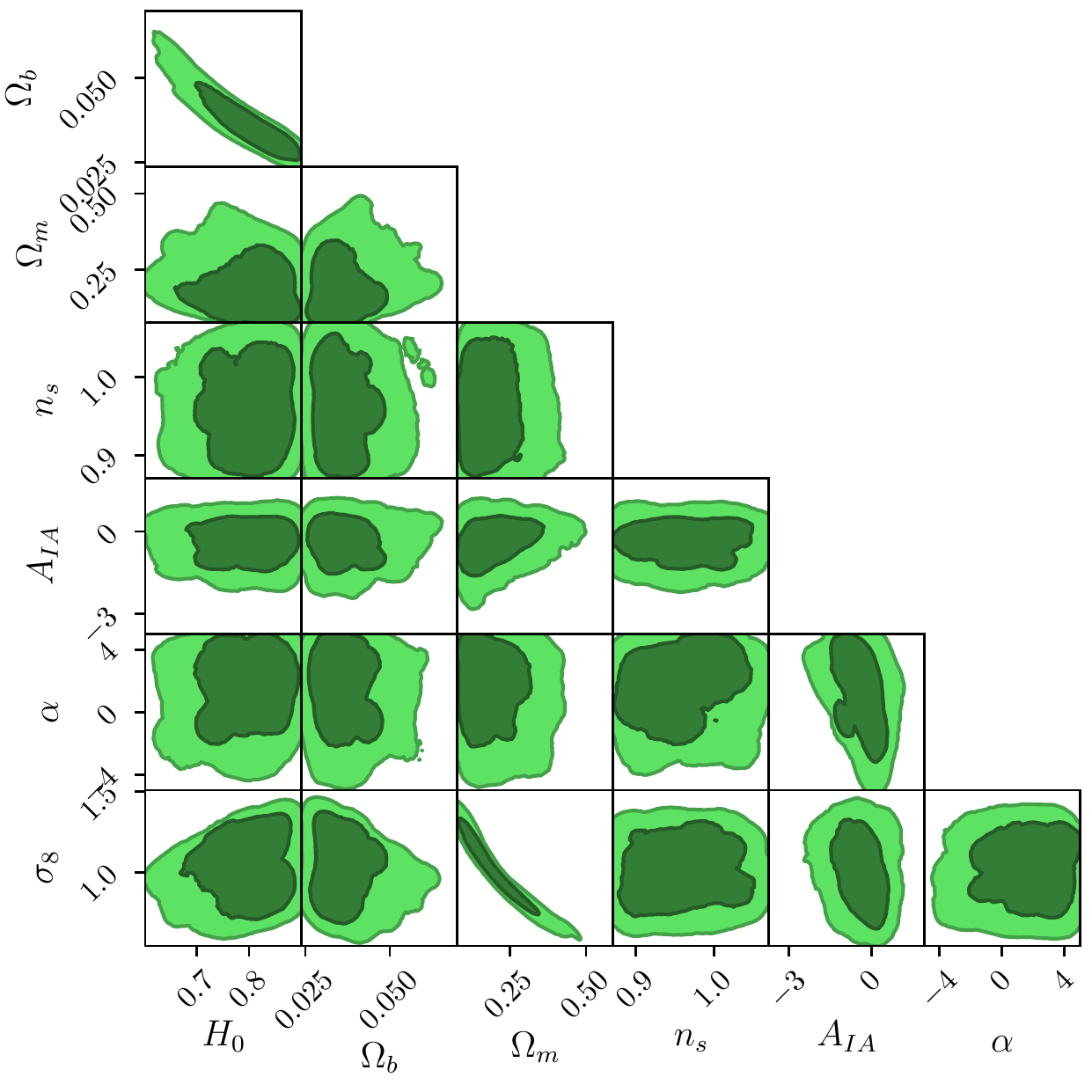}
    \caption{The $68 \%$ and $95 \%$ confidence regions for the $k$-cut analysis with a target $k$-cut of $1 \ h {\rm Mpc}^{-1}$. It is worth comparing with the HSCY1 fiducial analysis shown in Fig. 21 of~\cite{Hikage_2019}. Despite the extremely conservative choice of scale cuts, the majority of the information is preserved (see also Fig.~\ref{fig:test cut}). All contour plots appearing in this work were generated using Chainconsumer~\cite{Hinton2016} and the width of the kernel density estimator (KDE) bandpass is set to 1.5 unless explicitly stated otherwise.}
    \label{all_values}
\end{figure*}

 The resulting HSCY1 $k$-cut parameter constraints in the $\Omega_m - \sigma_8$ and $\Omega_m - S_8$ planes are shown Fig.~\ref{fig:s8 compare} and the full constraints are shown in Fig.~\ref{all_values}. We find $S_8 =  0.789 ^{+0.039}_{-0.022}$ in the $k$-cut analysis. Despite taking a very conservative choice of scale cut, the symmetric error on $S_8$ increases by just $3 \%$ compared to the HSCY1 constraints.

\begin{table}
    \begin{tabular}{ || c |  c || }
    \hline
    Parameter  & Prior \\
    \hline
    \hline
     $\Omega_m h^2$ & flat[0.03, 0.7] \\
     $\Omega_b h^2$ & flat[0.019, 0.026] \\
    $H_0$ & flat[0.6, 0.9] \\
     $\ln(10^{10}A_s)$ & flat[1.6, 6] \\
     $n_s$ & flat[0.87, 1.07] \\
    $\sum{m_\nu}$ & fixed(0) \\
    $w$ & fixed(-1) \\
    
    \hline
    
     $A_{\rm IA}$ & flat[-5, 5] \\
     $\alpha $ & flat[-5, 5] \\

    \hline
     $\Delta m^i$ & Gauss(0, 0.01) \\
     $100 m_{\rm sel}^1$ & 0.86 \\
    $100 m_{\rm sel}^2$ & 0.99 \\
    $100 m_{\rm sel}^3$ & 0.91 \\
    $100 m_{\rm sel}^4$ & 0.91 \\
    
    $100 m_\mathcal{R}^1$ & 0.0 \\
    $100 m_\mathcal{R}^2$ & 0.0 \\
    $100 m_\mathcal{R}^3$ & 1.5 \\
    $100 m_\mathcal{R}^4$ & 3.0 \\
     
    $\Delta z^1$ & Gauss(0, 0.0285) \\
    $\Delta z^2$ & Gauss(0, 0.0135) \\
    $\Delta z^3$ & Gauss(0, 0.0383) \\
    $\Delta z^4$ & Gauss(0, 0.0376) \\
    
    \hline
    \end{tabular}
    
    \caption{Prior range used in the fiducial $k$-cut $\Lambda$CDM analysis. The parameters are arranged into cosmological parameters (top), intrinsic alignment parameters (middle) and nuisance parameters (bottom). The prior ranges are the same as in the fidicial HSCY1 analysis~\cite{Hikage_2019}, except we assign a multiplicative bias to each tomographic bin as there is no \textit{a priori} reason to expect the multiplicative bias in each bin to be the same.}
    \label{tab:2}
\end{table}

\subsection{$f(R)$ results}
We perform a $k$-cut analysis to constrain Hu-Sawicki $f(R)$ gravity using the nonlinear emulator developed in~\cite{Ramachandra:2020lue} to generate the nonlinear power spectrum. As in the $\Lambda$CDM case we take a target $k_{\rm cut} = 1 \ h {\rm Mpc}^{-1}$ corresponding to the scale where the emulator's accuracy exceeds the 5\% level~\cite{Ramachandra:2020lue}. We take our priors to cover the region of parameter space where the emulator was trained (see Table~\ref{tab:fr_prior}), while for the nuisance and intrinsic alignment parameters we take the same priors as in the $\Lambda$CDM case. 
\par
\begin{figure} 
    \centering
    \includegraphics[width = \columnwidth]{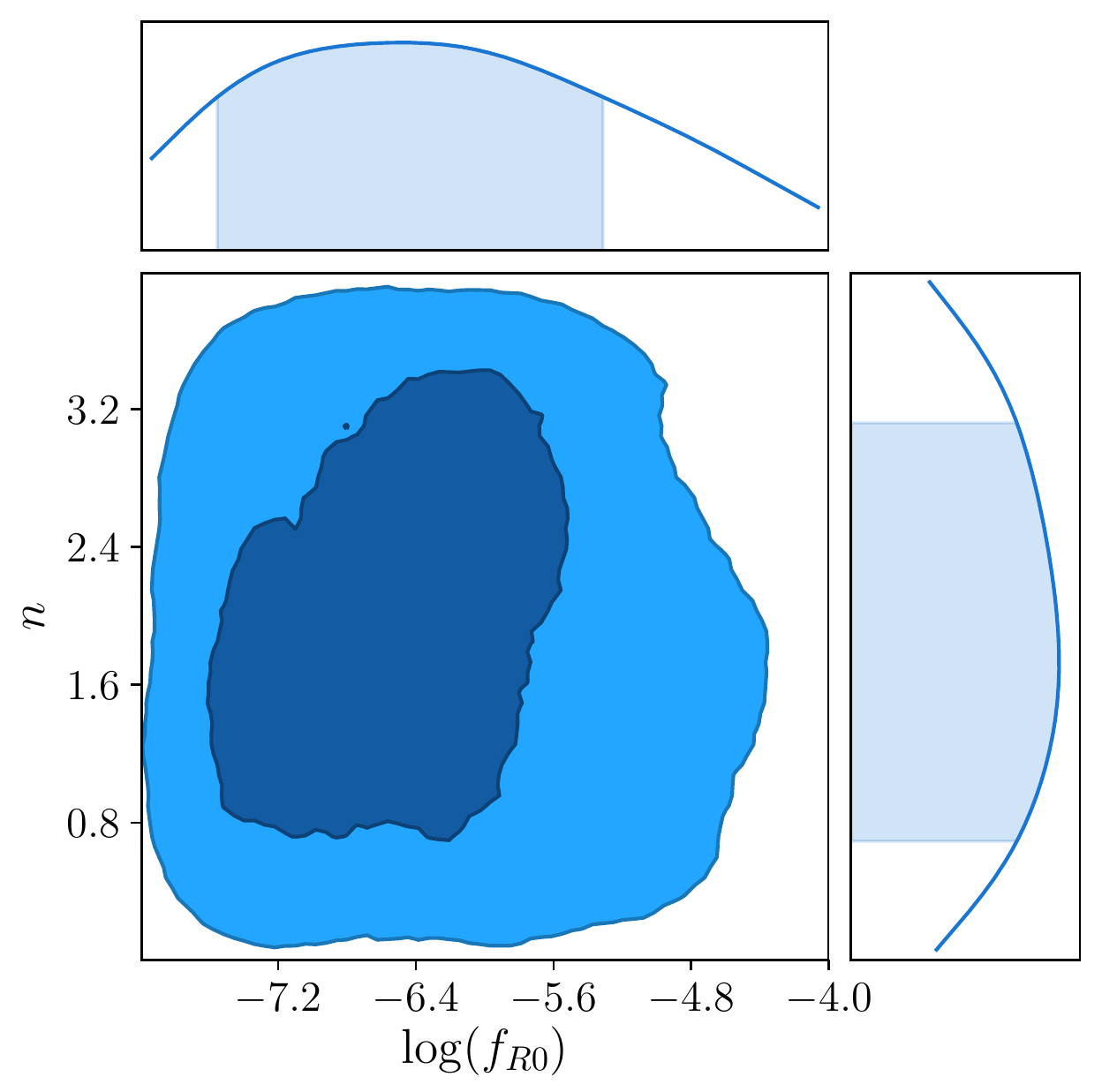}
    \caption{$k$-cut constraints on the Hu-Sawicki modified gravity parameters $n$ and $\log (f_{R_0})$. The edge of the plot indicates the prior boundary. We find  $\log (f_{R_0}) = -6.38^{+0.94}_{-1.41}$ and $ n = 1.8^{+1.1}_{-1.5}$. The constraints are nearly completely prior dominated although the posterior has slightly more weight at lower values of $\log(f_{R_0})$. }
    \label{fig:fr_result}
\end{figure}

 We show the results of the 2 Hu-Sawicki parameters in Fig.~\ref{fig:fr_result}. Our constraints, $\log (f_{R_0}) = -6.38^{+0.94}_{-1.41}$ and $ n = 1.8^{+1.1}_{-1.5}$, are almost completely prior dominated. 

\par We also repeat the likelihood analysis fixing $n = 1$ as in the analysis of~\cite{Troster:2020kai}. In this case we find $\mathrm{log}(f_{R_{0}}) = -6.67^{+1.37}_{-0.88}$. These results are consistent with those in~\cite{Troster:2020kai} which found that $\log (f_{R_0})$ could not be constrained within a uniform prior  $-8 < \log (f_{R_0}) <-2$ using  Kilo-Degree Survey (KiDS) data. External constraints rule out $f_{R_0} > 10^{-6}$~\cite{Lombriser:2014dua, Vikram:2014uza, Desmond:2020gzn}, so we must wait for next generation weak lensing data sets to place tighter constraints.

Although the region of parameter space containing $\Lambda$CDM is 
not contained in the prior, i.e., $(f_{R_0} = 0)$, there is a very weak preference for small values of $\log (f_{R_0})$ consistent with $\Lambda$CDM.
 
\section{Conclusion} \label{sec:conclusion}

In this work, we have performed the first ever $k$-cut cosmic shear analysis to constrain $\Lambda$CDM and Hu-Sawicki $f(R)$ gravity. The $k$-cut method is ideally suited to test theories of modified gravity for which we do not have accurate models for the nonlinear power spectrum as deep into the nonlinear regime as for the $\Lambda$CDM case. We choose a target $k$-cut of  $k_{\rm cut} = 1 \ h {\rm Mpc}^{-1}$ for both analyses as this is the scale where the nonlinear $f(R)$ emulator becomes inaccurate. 
\par By taking a simple linear transformation of the original data vector, the BNT transform tightens the correspondence between angular and physical scales. Angular scale cuts can then be used to cut sensitivity to poorly modeled small scales. The method comes with virtually no additional computational overhead and the strength of the $\ell - k$ relationship will only tighten as the number of tomographic bins is increased and photometric redshift error is reduced with future datasets.
\par The $k$-cut cosmic shear method comes with several caveats. The BNT-transformed kernels although narrow in $z$-space still have some width, so the relationship between $\ell$ and $k$ is not exact. Furthermore the BNT transformation itself depends on the tomographic selection function $n^j(z)$, so the efficacy of the transformation at separating physical scales could in principle be compromised by photometric redshift errors.  
\par To ensure that the method removes sensitivity to small scales as intended, we fix the cosmological parameters and constrain a single free parameter, $\mathcal{A}_p$, which controls the amplitude of the matter power spectrum above the target scale cut in $k$-space. We find the error on this amplitude increase by a factor of 5.5 in the $k$-cut case demonstrating that the sensitivity to these scales is significantly reduced as intended. We recommend this test is performed in all future $k$-cut cosmic shear and $k$-cut $3 \times 2$ point~\cite{Taylor:2020imc} studies.
\par In the $\Lambda$CDM case, we find $S_8 = 0.789 ^{+0.039}_{-0.022}$. Our choice of scale cut ensures the constraints are robust to baryonic physics model uncertainties. Despite a conservative choice of scale cut, the symmetric error on $S_8$ increases by just $3 \%$ relative to the HSCY1 angular power spectrum analysis~\cite{Hikage_2019}.
\par To constrain $f(R)$ gravity, we used the nonlinear power spectrum emulator developed in~\cite{Ramachandra:2020lue}. After marginalizing over all other parameters we find $\log (f_{R_0}) = -6.38^{+0.94}_{-1.41}$ and $ n = 1.8^{+1.1}_{-1.5}$. 
\par Although we find weak lensing constraints on $f(R)$ gravity are prior dominated, this will soon change as data from Euclid, Roman and the Rubin Observatory arrives over the next decade. The $k$-cut method, implemented for the first time in this paper, is a promising approach to test modified gravity into the nonlinear regime while avoiding model bias.

\section{Acknowledgements}
 L.V. acknowledges support from a Caltech Summer Undergraduate Research Fellowship (SURF) and thanks Alice and Edward Stone for providing the funding. P.L.T. acknowledges support for this work from a NASA Postdoctoral Program Fellowship. Part of the research was carried out at the Jet Propulsion Laboratory, California Institute of Technology, under a contract with the National Aeronautics and Space Administration. The work of G.V. is financially supported by NSF grant No. AST-1813694. N.S.R.'s work at Argonne National Laboratory was supported under the US Department of Energy contract No. DE-AC02-06CH11357. The authors are indebted to Chiaki Hikage for providing the HSCY1 extended scale data vector and covariance matrix and the tomographic photometric redshift distributions. We would also like to Alex Hall for pointing out a more streamlined approach to compute the BNT covariance matrix and Vincenzo Cardone for useful discussions.

\bibliographystyle{apsrev4-1.bst}
\bibliography{ref.bib}

\appendix
\section{$k$-cut $3 \times 2$ Point Covariance} \label{sec:cov2}
The generalization of the $k$-cut method to $3 \times 2$ point statistics is developed in~\cite{Taylor:2020imc}. In this case the expression for the data vector covariance in Eqn.~\ref{eqn:cov} must be extended to the full $3 \times 2$ point BNT-transformed data vector. 
\par If $\widehat C_{s_1,s_2}^{ef,gh} \ell, \ell')$ is an estimate of the covariance between $C_{s_1}^{ef}(\ell)$ and $C_{s_2}^{gh}(\ell)$, where $s_i \in \{  {\rm L}, {\rm G}, {\rm GGL}   \}$ labels whether the power spectra correspond to cosmic shear, galaxy clustering or galaxy-galaxy lensing respectively, then the BNT-transformed covariance is given by
\begin{equation}
\begin{aligned}
    \widehat C_{s_1,s_2,{\rm BNT}}^{ab,cd}( \ell, \ell') = X_{s_1}^{aebf} X_{s_2}^{cgdh} \\ \times \widehat C_{s_1,s_2}^{ef,gh}( \ell, \ell'),
\end{aligned}
\end{equation}
where 

  \begin{equation}
    X_{s}^{aebf}=
    \begin{cases}
      M^{ae}M^{bf}, & \text{if}\ s={\rm L} \\
      \delta^{ae}\delta^{bf}, & \text{if}\ s={\rm G} \\
      \delta^{ae}M^{bf}, & \text{if}\ s={\rm GGL}, \\
    \end{cases}
  \end{equation}
$M^{ab}$ denotes the BNT matrix, $\delta^{ae}$ denotes the Kronecker delta and repeated indices are summed over.
\end{document}